\documentclass[twocolumn]{aastex62}

\usepackage{amsmath,amssymb}
\usepackage{epstopdf}
\usepackage{ulem}
\usepackage{comment}

\submitjournal{ApJ}

\shorttitle{Magnetic reconnection at SNR shocks}
\shortauthors{Bohdan et al.}

\newcommand{\revtwo}{\textcolor{black}}
\newcommand{\rev}{\textcolor{black}}

\newcommand{\mpo}{\textcolor{black}}
\newcommand{\ab}{\textcolor{black}}
\newcommand{\jn}{\textcolor{black}}

\def\ee{\end{equation}}
\def\be{\begin{equation}}
\def\l{\left}
\def\r{\right}
\def\pa{\partial}

\newcommand{\thbn}{\Theta_{\rm Bn}}

\newcommand{\ompe}{\omega_\mathrm{pe}}
\newcommand{\ompi}{\omega_\mathrm{pi}}
\newcommand{\omce}{\Omega_\mathrm{e}}
\newcommand{\omci}{\Omega_\mathrm{i}}

\newcommand{\ms}{M_\mathrm{s}}
\newcommand{\ma}{M_\mathrm{A}}
\newcommand{\mi}{m_\mathrm{i}}
\newcommand{\me}{m_\mathrm{e}}
\newcommand{\lse}{\lambda_\mathrm{se}}
\newcommand{\lsi}{\lambda_\mathrm{si}}
\newcommand{\vsh}{v_\mathrm{sh}}

\begin{document}

\title{Kinetic simulations of nonrelativistic perpendicular shocks of young supernova remnants. III. Magnetic reconnection.}

\correspondingauthor{Artem Bohdan}
\email{artem.bohdan@desy.de}

\author[0000-0002-5680-0766]{Artem Bohdan}
\affil{DESY, 15738 Zeuthen, Germany}

\author{Martin Pohl}
\affil{DESY, 15738 Zeuthen, Germany}
\affil{Institute of Physics and Astronomy, University of Potsdam, 14476 Potsdam, Germany}

\author{Jacek Niemiec}
\affil{Institute of Nuclear Physics Polish Academy of Sciences, PL-31342 Krakow, Poland}

\author{Sergei Vafin}
\affil{Institute of Physics and Astronomy, University of Potsdam, 14476 Potsdam, Germany}

\author{Yosuke Matsumoto}
\affil{Department of Physics, Chiba University, 1-33 Yayoi-cho, Inage-ku, Chiba 263-8522, Japan}

\author{Takanobu Amano}
\affil{Department of Earth and Planetary Science, the University of Tokyo, 7-3-1 Hongo, Bunkyo-ku, Tokyo 113-0033, Japan}

\author{Masahiro Hoshino}
\affil{Department of Earth and Planetary Science, the University of Tokyo, 7-3-1 Hongo, Bunkyo-ku, Tokyo 113-0033, Japan}

\begin{abstract}
Fully kinetic two-dimensional particle-in-cell simulations are used to study electron acceleration at high-Mach-number nonrelativistic perpendicular shocks. SNR shocks are mediated by the Weibel instability which is excited because of an interaction between shock-reflected and upstream ions. Nonlinear evolution of the Weibel instability leads to the formation of current sheets. At the turbulent shock ramp the current sheets decay through magnetic reconnection. The number of reconnection sites strongly depends on the ion-to-electron mass ratio and the Alfv\'enic Mach number of the simulated shock. Electron acceleration is observed at locations where magnetic reconnection operates. For the highest mass ratios almost all electrons are involved in magnetic reconnection, which makes the magnetic reconnection the dominant acceleration process for electrons at these shocks. We discuss the relevance of our results for 3D systems with realistic ion-to-electron mass ratio.
\end{abstract}


\section{Introduction}\label{introduction}

Deciphering the acceleration mechanisms of charged particles in space is of great interest and high actuality in astroplasma physics. The interaction of supernova ejecta with the interstellar medium results in shocks which are often associated with nonthermal radiation.
It is widely assumed that relativistic particles responsible for this emission are produced through diffusive shock acceleration (DSA, e.g., \cite{1977ICRC...11..132A,1983RPPh...46..973D,1987PhR...154....1B}). DSA relies on multiple interactions of particles with the shock front while \mpo{they are} confined near the shock by magnetic turbulence.
A critical ingredient and the main unsolved problem of DSA is the particle injection. Particles need to see the shock as a sharp discontinuity in the plasma flow to cross it unaffected, \mpo{and so} DSA works for high energy particles only. The shock has a finite width though that is commensurate with the gyroradius of the incoming ions. \mpo{The electron injection problem is much harder than that of ions on account of the} small electron mass. \mpo{As electrons and ions are not in equilibrium immediately behind a collisionless shock, an electron requires a considerably larger factor of energy increase to render its gyroradius comparable to that of ions} or the shock width. 
The problem of electron injection at \jn{supernova remnant (SNR)} shocks has remained unresolved for many years. It requires extensive studies because it determines the level of cosmic-ray feedback and hence the nonlinearity of the system. Also, in most cases it determines the X-ray and the gamma-ray luminosity of SNRs~\citep{2018A&A...618A.155S,2019A&A...627A.166B,2019arXiv190908484B}.

Here we continue our study of electron acceleration processes for conditions at young SNR shock waves.
\ab{Observational data and numerical simulations cannot still clarify which magnetic-field configuration in SNR shocks is preferable for efficient electron acceleration. \mpo{Strong electron acceleration has been reported \jn{from} 2D simulations of quasi-parallel mildly relativistic shocks \citep{2019MNRAS.485.5105C} and \jn{from} 1D simulations of quasi-perpendicular shocks \citep{2019arXiv190807890X}.} However, SNR shocks are nonrelativistic and 1D simulations are not capable to reproduce fully correct shock physics. 
Therefore we study nonrelativistic shocks with perpendicular magnetic-field configuration $\theta_{Bn}=90^o$ using 2D simulations.}

Young SNR shocks have nonrelativistic propagation speeds and \jn{they} are characterized by high sonic and Alfv\'enic Mach numbers, $\ma \approx 200$. In the high-Mach-number regime a portion of upstream ions is reflected back upstream by the shock potential~\citep{Marshall1955}. Reflected ions interact with upcoming plasma and drive various instabilities in the shock transition. The shock transition is subdivided into an upstream, a foot, a ramp, an overshoot, and the downstream region. The undisturbed plasma is contained upstream and the shocked plasma is found downstream. 
\jn{In a quasi-perpendicular shock the}
interaction of reflected ions with upstream electrons leads the excitation of electrostatic Buneman waves \citep{1958PhRvL...1....8B} at the leading edge of the shock foot, and the two-stream ion-ion Weibel instability \citep{1959PhFl....2..337F} operates deeper in the shock foot. Ion reflection occurs at the shock ramp, and a strong rise of a plasma density is observed here that reaches its maximum at the shock overshoot.

The Buneman instability accelerates electrons via shock surfing acceleration \citep[SSA][]{2000ApJ...543L..67S,2002ApJ...572..880H} when they coherently interact with electrostatic waves. The Weibel instability is responsible for electron energization via magnetic reconnection~\citep{Matsumoto2015}, stochastic Fermi-like acceleration~\citep{Bohdan2017}, and stochastic shock drift acceleration~\citep{Matsumoto2017}.
Since these instabilities partially operate on very small scales between the electron inertial length and the ion skin \ab{depth}, and also because electron acceleration is the subject of investigation, the appropriate numerical tool are full Particle-in-Cell (PIC) simulations that treat both ions and electrons as particles.

This paper is the third in a series of works focusing on the analysis of high-Mach-number perpendicular shock \jn{with the method of} PIC simulations.
Previously \citep[hereafter Paper I]{Bohdan2019a} we discussed SSA at the leading edge of the shock foot and \mpo{the dependence of its efficiency} on the Mach number, ion-to-electron mass ratio and the magnetic field configuration.  
The second part \citep[hereafter Paper II]{Bohdan2019b} is devoted to \mpo{the impact} of SSA on the downstream nonthermal electron population. We found that SSA \mpo{negligibly contributes} in systems with strictly perpendicular configurations and realistic mass ratio.
However, the SSA mechanism remains important at quasi-perpendicular shocks. Electrons pre-accelerated by SSA can be reflected back upstream via mirror reflection \citep{Amano2007} and \ab{may excite low frequency waves \citep{PhysRevLett.104.181102}.}


The main goal of this work is to explore magnetic reconnection at high-Mach-number shocks and to define its impact on the downstream electron nonthermal population.


\citet{Matsumoto2015} \mpo{demonstrated electron acceleration up to nonthermal energies at high-Mach-number perpendicular shocks using 2D simulations} with in-plane magnetic-field configuration ($\varphi=0^o$, \jn{see Sec.~\ref{sec:setup}}). They found that electrons can be accelerated via elastic collisions with a jet ejected from the X point or \jn{via} interactions with magnetic islands residing in the reconnection region. \mpo{The efficiency of this process is unknown and moreover might depend on the numerical parameters of the kinetic simulations that are meant to explore them.} \citet{Bohdan2017} reported that magnetic reconnection also occurs in simulations with $\varphi=45^o$.
Note that magnetic reconnection in the Weibel turbulence was
not observed in the 3D simulation of \cite{Matsumoto2017} possibly because \mpo{the Mach number of the shock was not large enough.} Here we also want to 
\jn{investigate} this point.

This study can also be important for the low-Mach-number regime. Our results can be rescaled to the conditions at the Earth bow shock where magnetic reconnection is observed in simulations \citep{2014PhPl...21f2308K,2019GeoRL..46.9352B} and detected by in-situ observations \citep{2019GeoRL..46.1177G,2019GeoRL..46..562W}.


The paper is organized as follows. We present a short description of simulations setup in Section~\ref{sec:setup}. The results are presented in Section~\ref{results}. Discussion and summary are in Section~\ref{summary}.


\section{Simulation Setup} \label{sec:setup}


   \begin{table*}[!t]
      \caption{Simulation Parameters}
         \label{table-param}
     $$ 
\begin{array}{p{0.07\linewidth}rcccccccc}
\hline
\hline
\noalign{\smallskip}
Runs  & \ \varphi\  & \ \mi/\me\  & \ \ompe/\omce\  &  \ma & \multicolumn{2}{c}{\ms} & \multicolumn{2}{c}{\beta_{\rm e}} \\
 &  & & & &\  ^*1 &\  ^*2 &\  ^*1 &\  ^*2  \\
\noalign{\smallskip}
\hline
\noalign{\smallskip}
A1, A2 & 0^o  & 50  & 12  & \ 22.6\  & \ 1104\  & 35 & 5 \cdot 10^{-4} & 0.5  \\
B1, B2 & 0^o   & 100 & 12  &   31.8  & 1550 & 49 & 5 \cdot 10^{-4} & 0.5 \\
C1, C2 & 0^o  & 100 &  17.3 &   46  & 2242 & 71 & 5 \cdot 10^{-4} & 0.5 \\
D1, D2 & 0^o  & 200 & 8.5  &   32  & 1550 & 49 & 5 \cdot 10^{-4} & 0.5 \\
E1, E2 & 0^o  & 200 & 12  &   44.9  & 2191 & 69 & 5 \cdot 10^{-4} & 0.5  \\
F1, F2 & 0^o  & 400 & 13  &  68.7  & 3353 & \ 106\ & \ 5 \cdot 10^{-4}\ & \ 0.5\ \\
\noalign{\smallskip}
\hline
G1, G2 & 45^o  & 100 & 12  &  31.8  & 1550 & 49 & 5 \cdot 10^{-4} & 0.5\\
\noalign{\smallskip}
\hline
\end{array}
     $$ 
\tablecomments{Parameters of simulation runs described in this paper. Listed are: the ion-to-electron mass ratio, $\mi/\me$, the plasma magnetization, $\ompe/\omce$, and the Alfv\'enic  and sonic Mach numbers, $\ma$ and $\ms$, the latter separately for the \emph{left} (runs *1) and the \emph{right} (runs *2) shocks. We also list the electron plasma beta, $\beta_{\rm e}$, for each simulated shock.
Runs A-F use the in-plane magnetic field configuration and runs G have $\varphi=45^o$.} 
   \end{table*}

We performed simulations \jn{in which shocks are initialized} by means of the flow-flow method, as in our previous works (Paper I, Paper II, and \citet{Bohdan2017}).
Collision of two counterstreaming electron-ion plasma slabs leads to the formation of two shocks separated by a contact discontinuity. Hereafter we refer to both the flows and the shocks as the \emph{left} (L) and the \emph{right} (R). The absolute values of the beams velocities are $v_{\rm L}=v_{\rm R}=v_{\rm 0}=0.2c$.
Properties of colliding flows are identical except the plasma temperature which differs by factor of $1000$. It results in different plasma beta (the ratio of the electron plasma pressure to the magnetic pressure), $\beta_{\rm e,L}=5 \cdot 10^{-4}$ for the left beam and $\beta_{\rm e,R}=0.5$ for the right beam.


The large scale magnetic field, carried by the upstream plasma, is perpendicular to the upstream plasma velocity and the shock normal, $\thbn=90^\circ$. The angle between magnetic field direction and the simulation plane is $\varphi$. Runs A-F assume the in-plane magnetic-field configuration, $\varphi=0^o$, and $\varphi=45^o$ is used in runs G.

The adiabatic index of the plasma is  $5/3$ for both magnetic-field configurations, and the shock compression ratio is $r=4$. The shock speeds in the \emph{simulation} frame and the \emph{upstream} frame are $0.067c$ and $0.263c$, respectively. The Alfv\'enic, $\ma=\vsh/v_{\rm A}$, and sonic, $\ms=\vsh/c_{\rm s}$, Mach numbers of the shocks are defined in the conventional \emph{upstream} reference frame. Here the Alfv\'en velocity is  $v_{\rm A}=B_{\rm 0}/\sqrt{\mu_{\rm 0}(N_e\me+N_i\mi)}$, where $\mu_{\rm 0}$ is the vacuum permeability, $N_i$ and $N_e$ are the ion and the electron number densities, and $B_0$ is the magnetic-field strength in the far-upstream region. 
The sound speed is defined as $c_{\rm s}=(\Gamma k_BT_{\rm i}/\mi)^{1/2}$, where $k_B$ is the Boltzmann constant and $T_{\rm i}$ is the ion temperature.


Weakly magnetized plasma is considered in our runs. The ratio of the electron plasma frequency, $\omega_{\rm pe}=\sqrt{e^2N_e/\epsilon_0\me}$, to the electron gyrofrequency, $\Omega_{\rm e}=eB_0/\me$, \jn{is} in the range $\omega_{\rm pe}/\Omega_{\rm e}=8.5-17.3$. Here, $e$ is the electron charge, and $\epsilon_0$ is the vacuum permittivity.




Spatial and temporal scales are given in terms of the upstream ion skin depth, $\lsi$, and the upstream ion Larmor frequency, $\Omega_{\rm i}$, respectively.
Common parameters for all simulations are the upstream electron skin depth $\lse=20\Delta$, the time-step $\delta t=1/40\,\omega_{\rm pe}^{-1}$ and the number density in the far-upstream region, which is 20 particle per cell for each species. $\Delta$ is the size of grid cells.
More detailed description of the simulations setup is given in Paper I.

Here we discuss results of seven large-scale numerical experiments (runs A--G), that feature in total fourteen simulated shocks. We refer to each of them as a separate simulation run and label the shocks in the left plasma ($\beta_{\rm e,L}=5 \cdot 10^{-4}$) with *1, and the right shocks with *2 ($\beta_{\rm e,R}=0.5$). The parameters of the simulation runs discussed in this paper are listed in Table~\ref{table-param}.

The simulations cover a wide range of ion-to-electron mass ratios and Alfv\'enic Mach numbers, which permits an investigation of the influence of these parameters on the electron acceleration efficiency and to scale our results to the realistic ion-to-electron mass ratio. 
In this work we \mpo{investigate the influence of magnetic reconnection on the nonthermal electron population downstream of the shocks and its scaling with the} ion-to-electron mass ratio and the shock Mach number.
\mpo{Some aspects of the simulation runs have already} been discussed in our previous papers, namely, runs B and G in \cite{Bohdan2017} and runs A--F in Papers I and II.

Here we use the relativistic electromagnetic 2D3V-adapted PIC code \mpo{THATMPI, developed from} TRISTAN \citep{Buneman1993} with MPI-based parallelization \citep{2008ApJ...684.1174N}. We use the method of \cite{2016ApJ...820...62W} to suppress the artificial
electromagnetic transient that results from the initial strong
electric-field gradient between the two plasma slabs, and optimization of the particle sorting \citep{10.1007/978-3-319-78024-5_15}.  \revtwo{A solver by \cite{Vay2008} is used to update particle positions. The second-order approximation of the particle shapes, so called triangular-shape-cloud, and \cite{Friedman1990} filter for electric and magnetic fields are used to suppress the numerical grid-Cherenkov short-wave radiation. The numerical model used in our simulations has been extensively tested and results of the most recent tests are in detail presented in the thesis of \cite{Bohdan_thesis}. For the time between injection of plasma and interaction with the shock (usually less than $0.5\Omega_i^{-1}$) the thermal energy of electrons and electromagnetic field energy are conserved to better than $1\,\%$. Therefore our setup is stable enough for shock simulations with electron plasma beta $\beta_e=5\cdot 10^{-4}$ and $\beta_e=0.5$.}

\section{Results} \label{results}

As mentioned in Section~\ref{introduction}, \mpo{spontaneous magnetic reconnection in the turbulent shock ramp was observed in 2D simulations with $\varphi=0^o$ and $\varphi=45^o$ magnetic-field configurations} \citep{Matsumoto2015,Bohdan2017}. In this section we discuss the effects of magnetic reconnection and \mpo{its} role in electron pre-acceleration. The discussion is based on the results of all runs presented in Table~\ref{table-param}. Simulation parameters differ in the magnetic-field orientation, the ion-to-electron mass ratio, and the Alfv\'enic Mach number.

\subsection{Properties and statistics of magnetic reconnection sites} \label{mag-recon-stat}

In this subsection we discuss the detailed structure of the shock foot and ramp where the Weibel instability operates. The Weibel-type filamentation instability arises from the interaction between shock reflected ions and upstream plasma ions. These filaments are associated with current filaments and filamentary magnetic fields \citep{2010ApJ...721..828K,Matsumoto2015,2016ApJ...820...62W,Bohdan2017}.

\begin{figure*}[htb]
\centering
\includegraphics[width=0.49\linewidth]{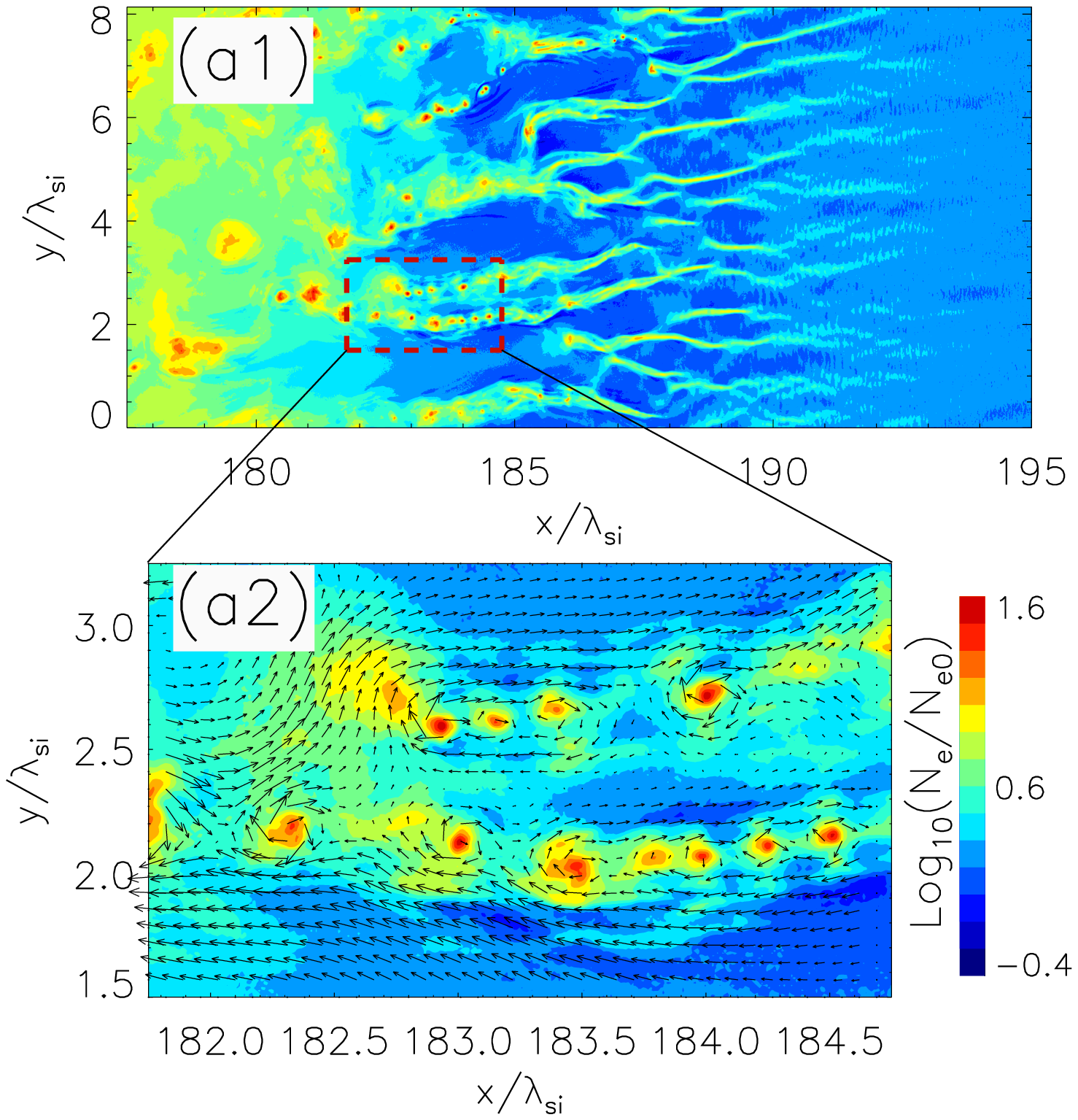}
\includegraphics[width=0.49\linewidth]{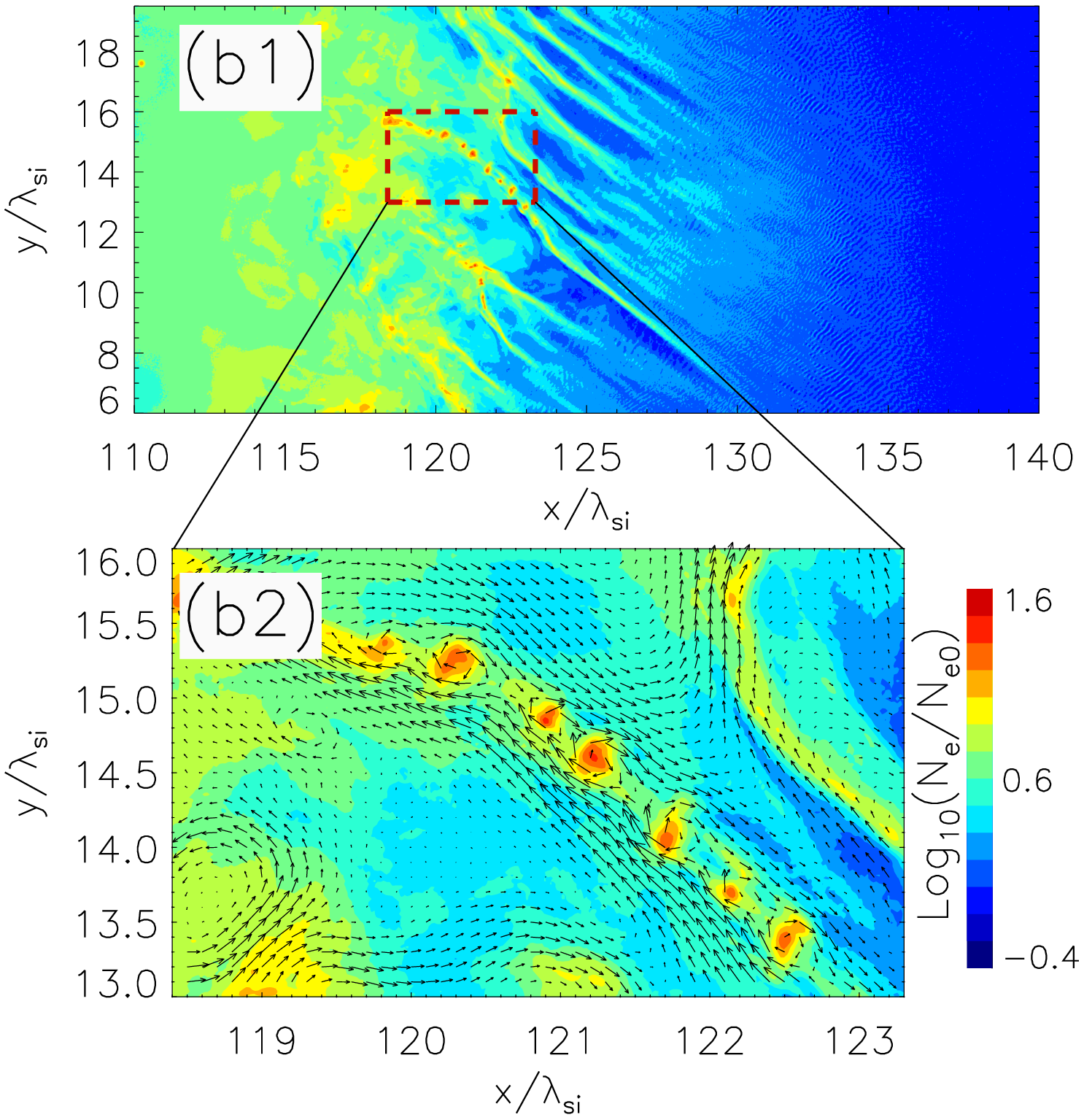}
\caption{\mpo{Left panels:} The shock ramp in run F2 ($\varphi=0^o$, panel (a1)) at time $t=4.3\omci^{-1}$. The region marked with dashed red lines in panel (a1) is shown enlarged in panel (a2). It harbors two chains of magnetic islands. 
 \mpo{Right panels:} The electron density distribution of the shock portion in run G2 ($\varphi=45^o$, panel (b1)) at time $t=3.3\Omega^{-1}$. The magnetic reconnection region is marked with dashed lines and panel (b2) is zoom-in of the reconnection region. The density is presented in a logarithmic scale and normalized to the upstream density. Arrows show the in-plane $xy$-component of the magnetic field.}
\label{reconnection-region}
\end{figure*}

   \begin{table}
         \caption{ Vortex parameters.}
         \label{table-vor-param}
\centering
\begin{tabular}{lccccc}
\hline
\hline
\noalign{\smallskip}
Run   & $\varphi$ & $m_i/m_e$ & VN & $\log_{10}$(AD$_\mathrm{sim}$) & Eq.~\ref{Weibel_nonlin}  \\
\noalign{\smallskip}
\hline
\noalign{\smallskip}
A1  &  $0^o$  & 50  & $0.1\pm0.1$  &  1.08 &  0.8 \\
A2  & $0^o$   & 50  & $0.11\pm0.1$ &  1.11 &  0.8  \\
B1  & $0^o$   & 100 & $0.61\pm0.34$ &  1.09 &  1.2  \\
B2  &  $0^o$  & 100 & $0.68\pm0.4$  &  1.08 &  1.2  \\
C1  & $0^o$   & 100 & $1.53\pm0.65$ &  1.01 &  1.7   \\
C2  &  $0^o$  & 100 & $1.36\pm0.76$ &  1.02 &  1.7  \\
D1  & $0^o$   & 200 & $0.89\pm0.48$ &  1.12 &  1.2  \\
D2  & $0^o$   & 200 & $0.87\pm0.37$ &  1.1  &  1.2  \\
E1  & $0^o$   & 200 & $1.97\pm0.95$ &  1.08 &  1.7  \\
E2  & $0^o$   & 200 & $2.34\pm0.82$ &  1.07 &  1.7  \\
F1  & $0^o$   & 400 & $5.03\pm1.6$  &  1.01 &  2.6  \\
F2  &  $0^o$  & 400 & $6.4\pm2.4$   &  1.03 &  2.6  \\
\noalign{\smallskip}
\hline
\noalign{\smallskip}
G1  & $45^o$  & 100  & $1.15\pm0.03$  &  0.86 &  1.2  \\
G2  & $45^o$  & 100  & $1.17\pm0.03$  &  0.87 &  1.2 \\
\noalign{\smallskip}
\hline
\end{tabular}
\smallskip 
\tablecomments{VN designates the number of vortices normalized by the transverse size of the simulation box; the errors are calculated as standard deviations of VN. AD$_\mathrm{sim}$ is the normalized average (through whole simulation) electron density inside magnetic vortices. The last column contains the left-hand side of equation~\ref{Weibel_nonlin}. }
   \end{table}

\begin{figure}[htb]
\centering
\includegraphics[width=0.95\linewidth]{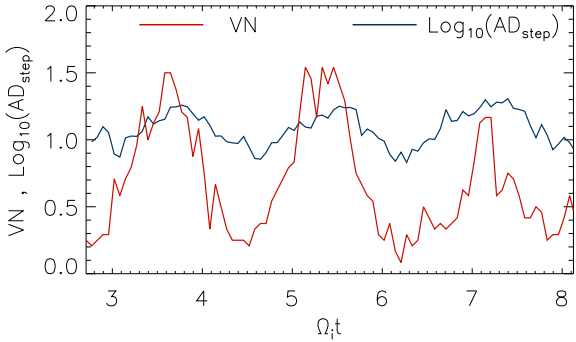}
\caption{Time evolution of the vortex number, VN (red line), and the logarithm of the normalized average electron density inside magnetic vortices, AD$_\mathrm{step}$ (blue line), \jn{for} run B2.}
\label{VGR-run-B1}
\end{figure}

\begin{figure}[htb]
\centering
\includegraphics[width=0.95\linewidth]{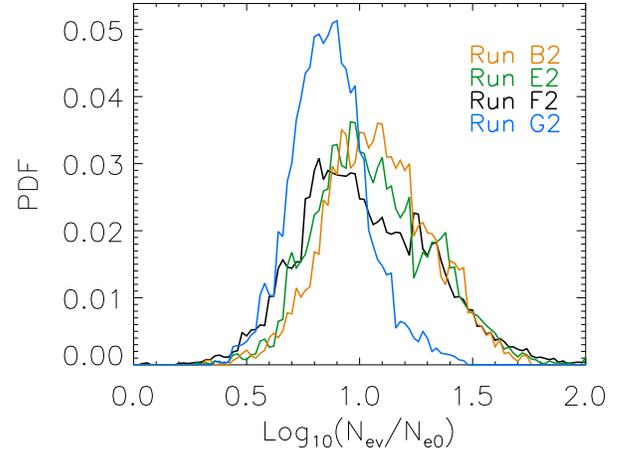}
\caption{Probability density of the logarithm of the electron density, $N_\mathrm{ev}$, inside magnetic vortices for runs B2 (yellow), E2 (green), F2 (black), and G2 (light blue).}
\label{vortex_distr}
\end{figure}

As shown in \cite{Matsumoto2015} for the in-plane magnetic-field configuration, magnetic filaments \mpo{in the shock ramp} can trigger spontaneous turbulent magnetic reconnection. We observe magnetic reconnection events not only for the in-plane (Fig.~\ref{reconnection-region}(a)), but also for the $\varphi=45^o$ configuration (Fig.~\ref{reconnection-region}(b)). 
In both cases, dense filaments represent 
current layers, 
which consists of a thin layer of dense plasma confined between two regions of oppositely directed magnetic field. Such a configuration is unstable and undergoes multiple magnetic reconnection forming X-points and magnetic islands.
It is natural that magnetic reconnection is observed in simulations with $\varphi=45^o$ configuration because the only difference in the structure of the shock ramp \jn{between the} in-plane and $\varphi=45^o$ configuration is the inclination of Weibel filaments, which depends on the direction of gyration of reflected ions in the ramp region, that, in turn, is defined by the orientation of the large-scale magnetic field \citep{Bohdan2017}. 

Figure~\ref{reconnection-region}(a1) displays a section of the foot/ramp region in run F2 at time $t=4.3\Omega^{-1}$. The leading edge of the foot with electrostatic Buneman waves is located at $x/\lsi \approx (200-207)$, and the overshoot is at $x/\lsi \approx 177$ (both regions are not shown in Fig.~\ref{reconnection-region}(a1)). The Weibel instability operates in the shock foot at $x/\lsi \approx (185-195)$. At the boundary between the foot and the ramp ($x/\lsi \approx 185$) the Weibel instability reaches a strongly nonlinear stage and magnetic reconnection occurs \jn{in the region} $x/\lsi \approx (177-185)$.
The existence of magnetic islands resulting from magnetic reconnection \jn{is} evident in the enlarged view in Figure~\ref{reconnection-region}a2. The density peaks are encircled by magnetic-field lines, which is a characteristic configuration for magnetic reconnection.
The magnetic-reconnection events can be identified as chains of magnetic islands separated by X-points, which result from nonlinear decay of the current sheets \citep{1963PhFl....6..459F}. 

The selected region \mpo{in Figure~\ref{reconnection-region}(a1) contains a variety of different structures}. Freshly formed dense filaments are at $x/\lsi \approx (187-190)$. Their separation scale is of the order of the ion inertia scale $\lsi$. Deeper in the ramp the filaments merge and undergo  magnetic reconnection at $x/\lsi \approx (181-187)$. At $x/\lsi \approx (177-181)$ dense single magnetic islands remain after magnetic-island coalescence. Note that shock self-reformation can strongly change the shock structure. In run F the extension of the Weibel instability regions varies in the range $L_\mathrm{ramp}/\lsi=5-20$ during one cycle of reformation, and favorable conditions for Weibel modes and magnetic reconnection exist only in the shock reformation phases with large filament extension. 


To quantify the effect of magnetic reconnection and compare different simulation runs we introduce the notion of magnetic vortices (or islands), their observed number, and the electron density inside the magnetic vortex. \mpo{A magnetic vortex is defined as a local maximum of the z-component of the vector potential that represents the in-plane magnetic field}.

The vortex number, VN, is defined as the number of magnetic vortices observed in the shock region at a given time, normalized by the transverse size of the simulation box in units of the ion skin \jn{depth}, $\lsi$. \mpo{For all runs, the time-averaged VN} is listed in Table~\ref{table-vor-param}. The number of reconnection sites grows with the ion-to-electron mass ratio and the Alfv\'enic Mach number. Note that \mpo{for small} mass ratios not all magnetic filaments decay via magnetic reconnection. One can see in Figure~\ref{reconnection-region}(b1), presenting the shock region for run G2 with $m_i/m_e=100$, that only one filament undergoes magnetic reconnection, whereas in the run F2 with $m_i/m_e=400$ (Fig.~\ref{reconnection-region}(a1)) all filaments finally show reconnection.
\jn{At a given Alfv\'enic Mach number} VN is almost twice as large for runs with $\varphi=45^o$ as with in-plane magnetic field (compare VN for runs B and G). \mpo{We do not observe a systematic correlation of VN with the plasma beta.}


\mpo{Variations in VN are observed in all simulation runs. Their amplitude depends on the coherency of the shock self-reformation along the shock. The time evolution of VN for run B2 is shown in Figure~\ref{VGR-run-B1} and is representative for all simulation runs. VN varies in the range $0.2-1.5$ with an average value about $0.7$. The period and the phase of these variations coincide with the period and the phase of the shock self-reformation. When the flux of reflected ions is small, magnetic filaments are almost absent, and VN is low}. Magnetic vortices can also be formed by turbulent plasma motions in the shock. Thus, even in the absence of magnetic filaments VN is never zero. The maximum of VN is observed when filaments have the largest extension and efficiently undergo magnetic reconnection.

\mpo{We also analyzed the average electron density inside magnetic vortices. 
The logarithm of the average density (AD$_\mathrm{sim}$) of electrons inside magnetic islands, normalized by the upstream electron density, is listed in Table~\ref{table-vor-param} for all runs.} \ab{Here we use the average density for the whole simulation, AD$_\mathrm{sim}$, and the average density for the single time step,  AD$_\mathrm{step}$.} 
Simulations with in-plane magnetic field (runs A-F) provide similar AD$_\mathrm{sim}$ values. The slightly higher value of AD$_\mathrm{sim}$ for run D1 may be a statistical fluctuation due to the poor vortex statistics. Simulations with $\varphi=45^o$ (runs G) yield smaller AD$_\mathrm{sim}$. The time evolution of AD$_\mathrm{step}$ in run B2 (blue line in Fig.~\ref{VGR-run-B1}) shows that \mpo{for very low VN, when vortices are generated by plasma turbulence or are the remainder of vortex coalescence, AD$_\mathrm{step}$ is small. This suggests that magnetic vortices generated via filaments decay are denser than vortices generated by plasma turbulence, at least before their coalescence is over.} 

The time-averaged probability density functions, PDFs, of the electron density \ab{inside magnetic vortices ($N_\mathrm{ev}$)} is presented in Figure~\ref{vortex_distr} for runs B2, E2, F2 and G2, \mpo{for which the vortex statistics is good}. 
The PDF for run G2 strongly differs from other runs with in-plane magnetic field, \mpo{that show similar distributions}. Filament decay in run G2 generates vortices with smaller density, and in addition many vortices arise \mpo{from magnetic turbulence and not from the decay of magnetic filaments. }

\rev{The magnetic field inside magnetic vortices represents the guide field in current sheets. Its strength is similar in all runs and equals $\vert B_\mathrm{guide}\vert\lesssim 8\,\vert B_\mathrm{0}\vert$. 
The strength of reconnecting magnetic field, $B_\mathrm{MR}$, is commensurate with that of the Weibel-generated magnetic field, $B_\mathrm{W}$. A few percent of the ion upstream kinetic energy goes to the Weibel-generated magnetic field \citep{2008ApJ...681L..93K}, whose strength hence is 
\be
\vert B_\mathrm{W}\vert \approx 0.1 \ma \vert B_0 \vert ,
\label{BWeibel}
\ee
thus $\vert B_\mathrm{guide}\vert \lesssim 80\,\vert B_\mathrm{MR}\vert /\ma$.
The strength of reconnecting magnetic field in 2D plane,
$B_\mathrm{x,y}$, is comparable with that of the guide field component $B_\mathrm{z}$ in all simulation runs, but we also found a tendency that the guild field gets weaker than the reconnecting field in shocks with high Alfv\'enic Mach number.}

We conclude that the properties of the vortices are defined by the magnetic-field configuration of the simulation. 
Since realistic 3D shocks, especially the Weibel instability region, are well represented by 2D simulations with in-plane configuration \ab{\citep{Bohdan2017}}, the next three subsections are dedicated to runs A--F.

\subsection{Explanation of vortex number} \label{vgr-results}


\begin{figure}[!t]
\centering
\includegraphics[width=0.99\linewidth]{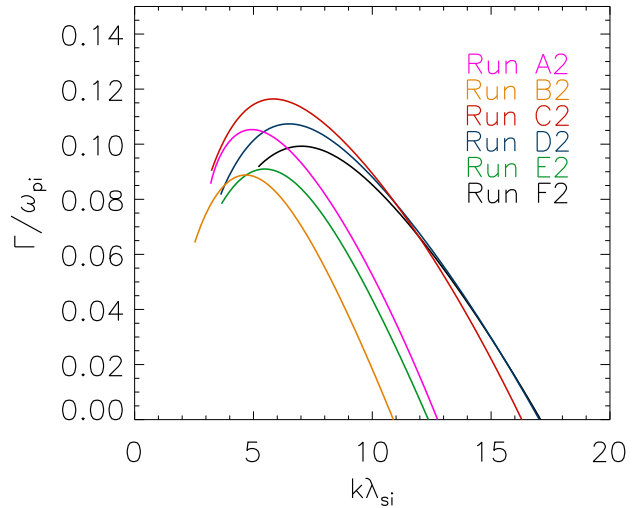}
\caption{The growth rate of the Weibel instability at the shock foot for runs A2 (magenta), B2 (yellow), C2 (red), D2 (blue), E2 (green), and F2 (black).}
\label{growsrate}
\end{figure}

It is important to understand \mpo{the efficiency scaling of the Weibel instability and electron preacceleration} by magnetic reconnection. At nonrelativistic shocks the Weibel instability results from the interaction of two relatively cold ion streams (shock reflected ions and upcoming upstream ions) \mpo{in a bath of} hot electrons. \cite{2010ApJ...721..828K} demonstrated consistence of their simulation results with an analytical description of the Weibel instability in the unmagnetized limit. We use a modification of this approach, \mpo{described in detail} in Appendix 1. 
For the analysis we chose regions in the shock foot at the stage of shock reformation when the number of the reflected ions is largest. The parameters of the plasma at this time are similar to those \ab{at the shock foot in simulation of \cite{2010ApJ...721..828K} (see Table 1, $x=2350$).}

\mpo{In Figure~\ref{growsrate} we present the growth rate of the ion Weibel instability as function of the wavenumber. The method of calculation is detailed in the Appendix \ref{appendix}. Once} the growth rate is normalized by the ion plasma frequency, $\ompi$, and the wavelength is normalized by the ion skin \ab{depth}, $\lsi$, we do not see a substantial \mpo{variation in the wavenumber of the most unstable mode and its growth rate.
The wavelength is in all cases} $\lambda_{WI} \approx \lambda_{si}$, and the growth rates is $\Gamma \approx 0.1 \ompi$. 
\mpo{This is not surprising because all simulations show} similar conditions at the shock foot.

\mpo{The Weibel instability forms dense current filaments that can be tearing-mode unstable.} Stability of the tearing mode is \mpo{controlled by the ratio of the perpendicular and the parallel} electron temperature \citep{1984PhFl...27.1198C}. We found that \mpo{$T_{\perp}/T_{\parallel} >1$ is always true inside Weibel filaments which renders} the tearing mode unstable.
We observe similar value of $T_{\perp}/T_{\parallel}$ inside Weibel filaments in all runs with in-plane magnetic field configuration. Figure~\ref{vortex_distr} shows that the plasma density \mpo{PDFs inside magnetic vortices are also similar. 
Therefore we can conclude that in runs A--F magnetic reconnection occurs at the same evolutionary stage of the Weibel instability.}

The distance plasma travels before the Weibel instability reaches its nonlinear stage is \mpo{
\be
L_\mathrm{W} \approx \frac{10}{\Gamma_\mathrm{max}} v_\mathrm{R} \approx 100 \ompi^{-1} v_0. 
\label{eq:lw}
\ee
\rev{where $\Gamma_\mathrm{max}$ is the growth rate of the most unstable mode} \revtwo{and the factor 10 is chosen ad hoc as an approximate boundary between linear and nonlinear stages of Weibel instability.}
If the shock foot length, $L_\mathrm{foot} \approx r_\mathrm{gi}= v_0/ \omci$, is longer than that, or}
\be
\label{Weibel_nonlin}
10^{-2} \ \frac{\ompi}{\omci} = 10^{-2}\sqrt{\frac{\mi}{\me}} \frac{\ompe}{\omce} > 1 \ ,
\ee 
the Weibel instability reaches its nonlinear stage, and magnetic reconnection \revtwo{can} occur. \mpo{Condition~\ref{Weibel_nonlin} indicates the likelihood and rate of magnetic reconnection in our simulations.} If equation~\ref{Weibel_nonlin} is not satisfied, as in run A, magnetic reconnection is almost absent. 
For simulations that exceed this limit by a similar margin, a comparable VN is observed (see Table~\ref{table-vor-param}).

\rev{Equation~\ref{Weibel_nonlin} defines only the average ability of Weibel filaments to decay via magnetic reconnection. The influence of local plasma conditions and shock self-reformation is strong for a small mass ratios \revtwo{($\mi/\me \lesssim 100$)}, because the Weibel instability barely reaches the nonlinear stage, and only some ``lucky" filaments undergo magnetic reconnection (see Fig.~\ref{reconnection-region}(b1)).}

In run F \mpo{essentially} all of the Weibel filaments undergo magnetic reconnection (see Fig.~\ref{reconnection-region}(a1)) which \mpo{suggests that a late nonlinear stage of the Weibel instability is reached.} For higher mass ratios VN can hence grow only because the Weibel filaments become longer. The length of the filaments scales with the shock thickness, and hence written in units of $\lsi$ it scales with the Alfv\'enic Mach number, $\ma$. As the average size of magnetic vortices is about $\lambda_{si}$, VN should grow linearly with $\ma$ at high mass ratio. We predict that \rev{$\mathrm{VN}_\mathrm{real} \approx \mathrm{VN}_\mathrm{Run F}*M_\mathrm{A,Run F}/M_\mathrm{A,real} \approx 12.5$  in case of realistic mass ratio and $M_\mathrm{A,real}\approx150$.}

\subsection{Acceleration processes due to magnetic reconnection} \label{acc-recon-results}

\begin{figure*}[!t]
\centering
\includegraphics[width=0.98\linewidth]{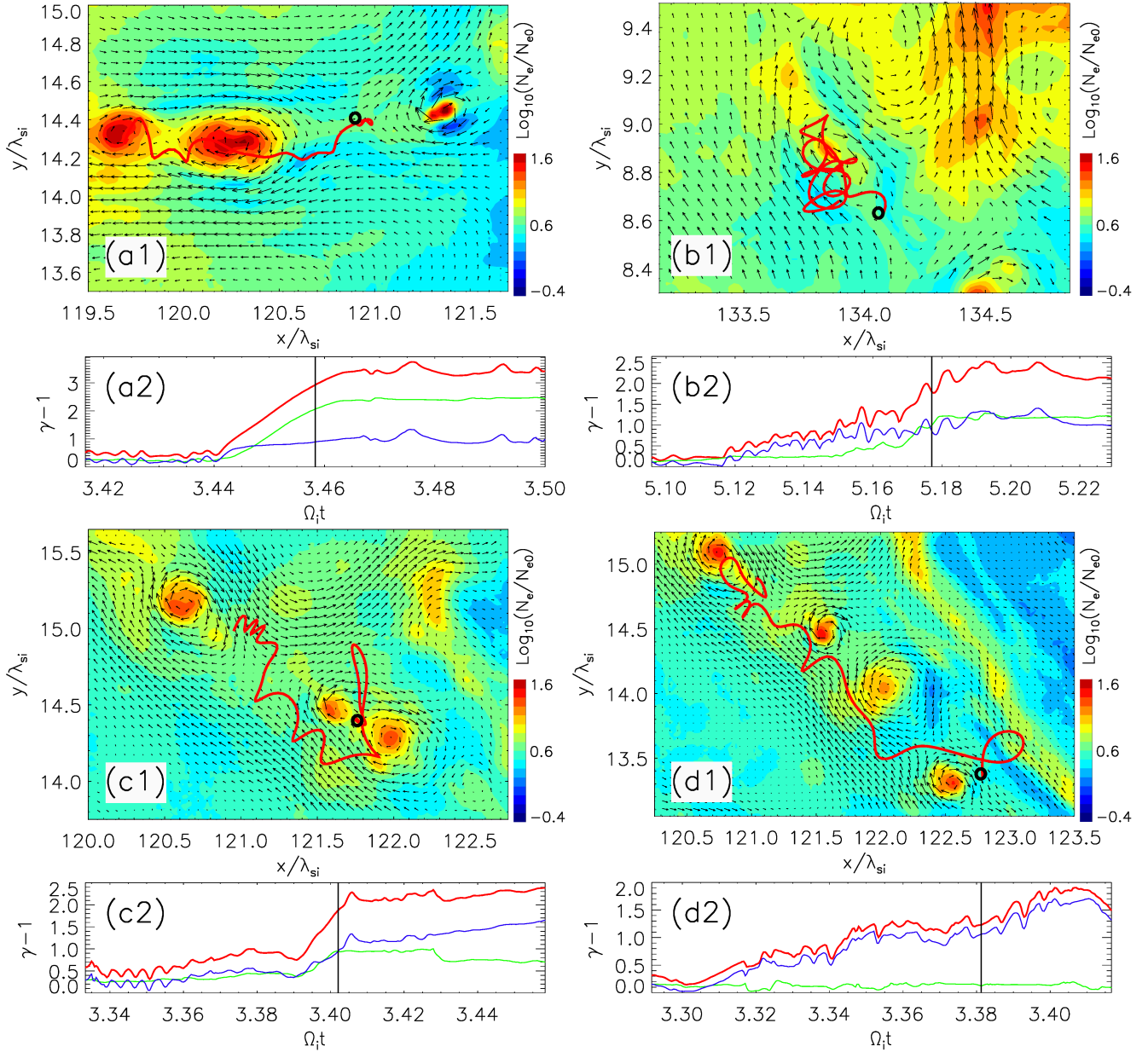}
\caption{Trajectories and energy evolution of four accelerated electrons: two for run B2 (panels (a*) and (b*)) and two for run \rev{G2} (panels (c*) and (d*)). Panels (*1): color -- normalized electron density in logarithmic scale \rev{at the specific time marked with vertical lines in panels (*2)}; arrows -- x-y magnetic-field lines; black circle -- position of particle at \rev{the same instance of time}; red line -- trajectory of the electron over \mpo{the last} $\Delta t=0.063\Omega_i^{-1}$. Panels (*2): Time evolution of the total kinetic energy of the electron (red line) and its parallel (green) and perpendicular (blue) components. 
}
\label{acc-particle}
\end{figure*}

It is well known that magnetic reconnection converts magnetic energy into thermal and kinetic particle energy \mpo{in a number of ways \citep[e.g.,][]{1965JGR....70.4219S,2006Natur.443..553D,2010JGRA..115.8223O,2001JGR...10625979H,2010ApJ...714..915O}.
Many studies of magnetic reconnection } \citep[e.g.,][]{2014PhPl...21i2304D,2015PhPl...22j0704D} use the so-called guiding center limit \citep{1963RvGSP...1..283N} for the identification of acceleration processes. This limit can be used if electromagnetic fields are constant on the scale of a particle's gyroradius in space and its gyroperiod in time. 

Characteristic spatial and temporal scales of magnetic reconnection are the ion skin \ab{depth}, $\lsi$, \ab{and the inverse ion gyrofrequency, $\Omega_\mathrm{i,MR}^{-1}$.} 
Therefore the guiding center limit can be used if 

\be
\label{gcl1}
\frac{\lsi}{r_\mathrm{ge,MR}}\gg 1 \ \ \textrm{and} \ \  \frac{\Omega_\mathrm{e,MR}}{\Omega_\mathrm{i,MR}} \gg 1 \ ,
\ee 
where $r_\mathrm{ge,MR}$ and $\Omega_\mathrm{e,MR}$ are the average gyroradius and the cyclotron frequency of an electron in the magnetic field of a reconnection region. 
\rev{Taking into account equation \ref{BWeibel}}
it follows that
\be
r_\mathrm{ge,MR}\approx\frac{10}{\ma}\frac{v_\mathrm{e}}{\omce} \ ,
\ee
\be
\Omega_\mathrm{e,MR} \approx \omce \frac{\ma}{10} \ 
\ee
and
\be
\Omega_\mathrm{i,MR} \approx \omci \frac{\ma}{10} \ ,
\label{eq:mr}
\ee
where $v_\mathrm{e}$ denotes the speed of the electron.
The temporal requirement is independent of the speed \ab{and it reduces to $\mi/\me \gg 1$ which is always satisfied in our simulations. The spatial requirement }
leads to the condition \revtwo{
\be
\label{gcl2}
0.1 \frac{\mi}{\me} \frac{\vsh}{\gamma_\mathrm{e} v_\mathrm{e}} \gg 1 \ .
\ee }
\rev{It is always satisfied for thermal electrons, if we assume that they are not in equilibrium with the ions, and so $v_\mathrm{e} \approx \vsh$. However, we are interested in high-energy electrons with $v_\mathrm{e} \approx c$ \revtwo{and $\gamma_\mathrm{e} \gg 1$}. For them equation 8 is not satisfied in run A and only marginally satisfied in all other runs.}
We conclude that the guiding-center limit can be used in our simulations with moderately reduced mass ratio (e.g., run F) and relatively high shock velocity, but for realistic shock SNR shock speeds, $\vsh \lesssim 0.02c$, the guiding center limit does not apply \rev{for high energy electrons} regardless of the mass ratio.


To perform an analysis of the acceleration processes we subdivide the particle energy into two parts, namely, the energy gained in the directions parallel and perpendicular to the local magnetic field. 
\mpo{In this way,} we can distinguish acceleration by the electric field along the magnetic field (the parallel component) and Fermi-like interaction with \mpo{moving} magnetic structures (the perpendicular component).

We shall now describe acceleration processes identified in our simulation runs. 
\mpo{In Figure~\ref{acc-particle} we present the trajectories of four representative electrons that reach nonthermal energies, $\gamma > 3$, well above the electron thermal energy downstream of the shock.} Electrons in panels (a1), (b1) and (c1), (d1) are selected from simulation runs B2 and G2, respectively. 

The first electron (Fig.~\ref{acc-particle}(a)) is accelerated by the $z$-component of the electric field at the X-point at $(x,y)/\lambda_{si}=(121,14.4)$. \rev{The guide magnetic field at the X-point is parallel to the $z$-axis and the local electric field, and it equals $3B_0$.}
During acceleration the electron stays in the vicinity of the X-point, and only the $z$-component of its momentum (perpendicular to the simulation plane) increases. As panel~\ref{acc-particle}(a2) shows, the rapid growth of the parallel component of the energy is observed at this stage, \rev{because the local magnetic and electric fields are parallel}. This is an example of the Speiser motion (see, e.g., \citealp{1965JGR....70.4219S}; \citealp{2001JGR...10625979H}).

The second electron (Fig.~\ref{acc-particle}(b)) is accelerated while it is captured by a magnetic vortex. 
\rev{The parallel component grows on account of acceleration by $E_z$, which is parallel to the local magnetic field inside the magnetic island. The perpendicular component is increased by adiabatic compression.}
We identify this acceleration process as \rev{a mixture of  ``island surfing" \citep{2010JGRA..115.8223O} (parallel component) and  adiabatic acceleration due to vortex contraction (perpendicular component).} 


The third electron (Fig.~\ref{acc-particle}(c)) experiences first-order Fermi acceleration by bouncing between merging magnetic islands. The electron undergoes several \rev{rapid} head-on collisions with magnetic ``walls" represented by magnetic vortices. \revtwo{During interactions with moving magnetic islands the electron is accelerated by a motional electric field, $\mathbf{E}=-\mathbf{v}\times\mathbf{B}$, thus this process contributes to the perpendicular energy gain component. The parallel energy growth occurs because of the $E_z$ field is present at the anti-X-point between merging islands. Combination of these two processes results in a continuous rise in energy during a short period of time at $\omci t \approx 3.40$.}

The fourth electron experiences second-order Fermi-like acceleration (Fig.~\ref{acc-particle}(d1)). Decay of Weibel filaments via magnetic reconnection produces a large number of magnetic vortices residing in the shock ramp and around the overshoot. Particles chaotically moving in these regions can be scattered by magnetic vortices, \mpo{some head-on and some tail-on.} This situation is similar to the second-order Fermi-like process discussed in \citet{Bohdan2017}. The trajectory of accelerated electron is presented in Figure~\ref{acc-particle}(d1). The total energy evolution is dominated by the perpendicular component, \revtwo{because of acceleration by motional electric field.}
The parallel energy component oscillates but remains constant on average. Regions with parallel electric and magnetic fields are small and rare, {and chaotic interaction with them provides} the almost steady parallel component.
This process is observed in all simulations with in-plane magnetic field \citep{Bohdan2017}, but here magnetic reconnection produces additional scattering centers that can increase its acceleration efficiency.


\mpo{The limited time available for magnetic reconnection in a self-reforming shock allows the identification of only a few acceleration processes. 
Simulation} studies dedicated to magnetic reconnection may cover $100\,\Omega_\mathrm{i,MR}^{-1}$ \citep[see, e.g.,][]{2010JGRA..115.8223O,2014PhPl...21i2304D} during which reconnection is steadily driven in a controlled manner. 
\mpo{The reconnection time observed in our shock simulations is 
\be
T_\mathrm{MR}\approx 0.5\, \omci^{-1} \approx 0.05\, \ma \Omega_\mathrm{i,MR}^{-1} \leq 3.5\, \Omega_{i,MR}^{-1} 
\ee
where we used Eq.~\ref{eq:mr}.}
Therefore, magnetic reconnection evolves from the filament formation to the emergence of a single magnetic vortex after island coalescence in less than $ 3.5 \Omega_\mathrm{i,MR}^{-1}$. 
Some processes require much more time than that, e.g., electron acceleration via contraction of magnetic islands \citep{2006Natur.443..553D}.
\mpo{Shocks with very high Alfv\'enic Mach number may thus} offer the conditions for larger variety of acceleration processes.

\rev{Note that ion heating/acceleration via magnetic reconnection is not observed. This is not unexpected for two reasons. First, only a few percent of the upstream ion kinetic energy goes to the Weibel-generated magnetic field \citep{2008ApJ...681L..93K} which limits the energy that can be transferred back to ions via magnetic reconnection. Then, the simulation is likely too short to cover ion-acceleration processes that operate on timescales of the ion gyro-period or longer.}


\subsection{Influence of magnetic reconnection on the downstream electron spectra} \label{down-sp-2-results}

   \begin{table*}
      \caption{Influence of magnetic reconnection.}
         \label{table-spectra2}
\centering
\begin{tabular}{lcccccccc}
\hline
\hline
\noalign{\smallskip}
Run   & $m_i/m_e$ & $M_A$ & $k_B T/m_ec^2$ & NTEF (\%)  & NTEF$_\mathrm{SSA}$ (\%)  & NTEF$_\mathrm{MR}$ (\%)  & $f_\mathrm{MR}$ (\%) \\
\noalign{\smallskip}
\hline
\noalign{\smallskip}
A2  &  50   & 22.6  & 0.09 &  0.28 & 0.15   & 0.1   & 10  \\
B2  &  100  & 31.8  & 0.18 &  0.55 & 0.05   & 0.4   & 26   \\
C2  &  100  & 46    & 0.22 &  0.36 & 0.05   & 0.13  & 38   \\
D2  &  200  & 32    & 0.3  &  0.7  & 0.025  & 0.55  & 38   \\
E2  &  200  & 44.6  & 0.37 &  0.56 & 0.025  & 0.5   & 43   \\
F2  &  400  & 68.7  & 0.73 &  0.57 & 0.0015 & 0.4   & 79   \\
\noalign{\smallskip}
\hline
\end{tabular}
\smallskip
\tablecomments{NTEF - nonthermal electron fraction. NTEF$_\mathrm{SSA}$ is the nonthermal electron fraction \mpo{arising from} SSA~\citep{Bohdan2019b}. NTEF$_\mathrm{MR}$ is the nonthermal electron fraction \mpo{attributable to magnetic reconnection. $f_\mathrm{MR}$ is defined as the number ratio of electrons involved in magnetic reconnection and all electrons that} passed through the shock.} 
   \end{table*}

\begin{figure}[htb]
\centering
\includegraphics[width=0.99\linewidth]{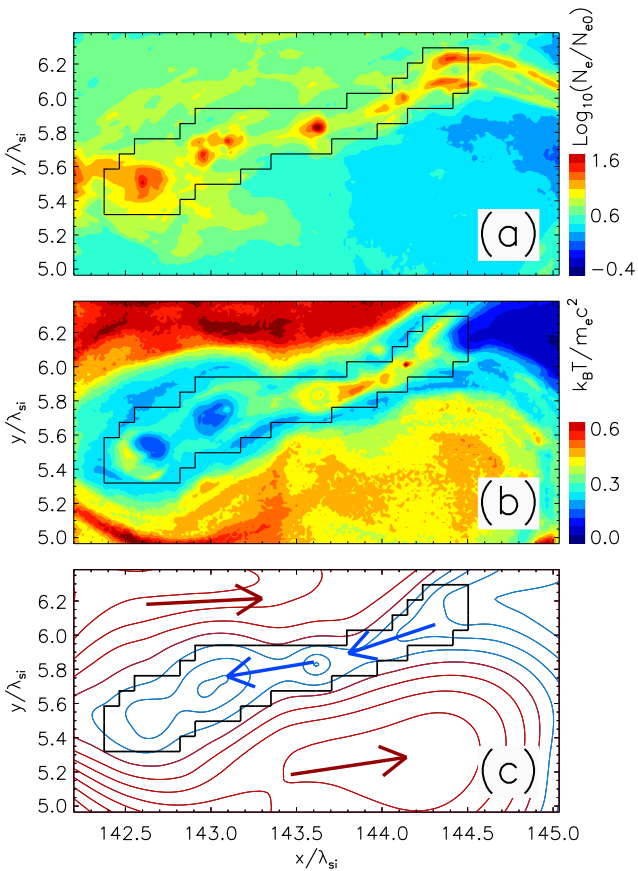}
\caption{Panel (a): \mpo{Electron density map. Panel (b): Electron temperature map. Panel (c): X-Y magnetic field lines. Arrows point in the direction of the bulk plasma motion. Red and blue colors refer to the plasma magnetically connected with the overshoot and the shock foot, respectively.} 
The reconnection region is marked by the black contour in all panels. \ab{Note that the shock upstream is on right side and the downstream region is on the left side.} }
\label{recon-site}
\end{figure}

In Paper II we discussed the influence of \mpo{SSA on the population of nonthermal electrons downstream of the shock, defined as the fraction of electrons in excess to a Maxwellian fit to the low-energy part of the downstream spectra. We found that SSA becomes less important  for higher mass ratios because electron heating in the shock transition is stronger then (see Table~\ref{table-spectra2}). The final} nonthermal-electron fraction (NTEF) remains roughly constant in all runs, about $0.5\%$. It \mpo{follows} that some other acceleration mechanism is responsible for the production of highly energetic electrons. 

In the shock foot two counterstreaming ion beams \mpo{provide the free energy to excite the ion Weibel instability~\citep{1959PhFl....2..337F,2008ApJ...681L..93K}. This instability deforms the magnetic field, forming dense filaments that during their evolution} may undergo magnetic reconnection.
Figure~\ref{recon-site} shows a section of the shock ramp \mpo{in which a Weibel filament undergoes magnetic reconnection. 
The plasma moves toward the shock overshoot and is concentrated} in dense filaments. \mpo{In Figure~\ref{recon-site}c the magnetic field lines are marked in blue and the bulk motion is indicated with} blue arrows. \mpo{Inside the filaments} the plasma is relatively cold and can be heated through magnetic reconnection or adiabatic compression.
\mpo{The regions between the Weibel filaments are hot and magnetically connected with the shock overshoot, indicated by the large red/yellow patches in panel (b) and the red magnetic field lines in panel (c). The bulk motion is toward the foot on the right of the panels.} \mpo{We conclude that the shock ramp is filled with a mixture} of cold plasma from the foot and hot plasma from the overshoot.

\mpo{Care must be exercised in defining magnetic reconnection regions to properly estimate the production efficiency of nonthermal electrons by reconnection. The vortex identification described in Section~\ref{mag-recon-stat} is one indicator.
The temperature distribution and the structure of the magnetic field can also be used \ab{to separate the cold foot plasma and the hot plasma from the shock overshoot}.
The black contour in Figure~\ref{recon-site} is based on all these indicators \ab{and selected manually}. It contains blue magnetic-field lines connected to the foot in panel (c) and the} relatively hot plasma inside the Weibel filament heated by magnetic reconnection (panel (b)). The contour defines the reconnection site and all particles inside are treated as particles involved in magnetic reconnection.

We \mpo{selected between six and eleven} reconnection regions for each shock propagating in warm plasmas ($\beta_{\rm e,L}=0.5$). \mpo{For these regions we calculate the number of electrons participating in reconnection and their contribution to the nonthermal electron population downstream of} the shock. \revtwo{Knowing how selected reconnection places contribute to the VN we can estimate the influence of magnetic reconnection on electron downstream spectra.
The results are listed in Table 3. $f_\mathrm{MR}$ is defined as the number ratio of electrons involved in magnetic reconnection and all electrons passing through the shock and can be calculated as
\be
f_\mathrm{MR}=\frac{N'_\mathrm{MR}}{\mathrm{VN}'_\mathrm{MR}} \cdot \frac{\mathrm{VN'}_{\omci}}{N'_{\omci}}
\ee
where $N'_\mathrm{MR}$ is the number of particles involved in magnetic reconnection at the selected places, $\mathrm{VN}'_\mathrm{MR}$ is VN generated at the selected reconnection regions, $\mathrm{VN'}_{\omci}$ is the number of vortices generated by the shock per inverse $\omci$, and $N'_{\omci}$ is the number of electron passing through the shock per inverse $\omci$. 
NTEF$_\mathrm{MR}$ is the nonthermal electron fraction produced at magnetic reconnection sites. It is defined as
\be
\mathrm{NTEF}_\mathrm{MR}=f_\mathrm{MR} \cdot \mathrm{NTEF}'_\mathrm{MR}
\ee
where $\mathrm{NTEF'}_\mathrm{MR}$ is the contribution to the NTEF by the selected reconnection regions.}
\mpo{As expected from the values of VN, $f_\mathrm{MR}$ is larger for runs with high Alfv\'enic Mach number and high mass ratio. In run F, $f_\mathrm{MR}$ reaches about $80\%$, and almost all Weibel filaments decay though magnetic reconnection.
We conclude that most nonthermal electrons are produced via magnetic reconnection. Its contribution, however, saturates at $\mathrm{NTEF_{MR}/NTEF}\approx0.8$ regardless of the value of $f_\mathrm{MR}$ (see values for runs D, E and F).} \revtwo{A possible explanation is that for high mass ratios the average energy of electrons involved in magnetic reconnection becomes comparable to or less than the downstream temperature. Therefore the importance of individual reconnection sites may decrease, but  this efficiency drop is balanced by a larger number of reconnection regions.}

One notices that $\mathrm{NTEF}_\mathrm{SSA}+\mathrm{NTEF}_\mathrm{MR}$ is always smaller than total fraction of nonthermal electrons, NTEF, and so an additional process may be at play. A good candidate is \jn{is the} chaotic interactions \jn{of particles} with magnetic turbulence at the shock ramp and \jn{in} the overshoot region.


Here we analysed magnetic reconnection only for the shocks in $\beta=0.5$ plasma. Shocks propagating in cold plasma ($\beta=5 \cdot 10^{-4}$) show a similar behaviour.

\section{Summary and discussion} \label{summary}

This paper is the third of a series investigating different aspects of electron acceleration at non-relativistic perpendicular shocks using \mpo{2D3V PIC simulations with different ion-to-electron mass ratios and Mach numbers.} 
Our previous studies \citep{Bohdan2019a,Bohdan2019b} indicated that SSA operating at the shock foot is not efficient enough to produce nonthermal electrons for \mpo{realistically large mass ratios even if the Alfv\'en Mach number is well above the trapping limit~\citep{Matsumoto2012} and the simulation design allows for a SSA efficiency as in 3D situations.
This paper investigates the influence of magnetic reconnection which results from the nonlinear decay of ion} Weibel filaments at the shock ramp.

Our main results are:

\begin{itemize}

\item Spontaneous turbulent magnetic reconnection in the shock transition is observed in the in-plane and $\varphi=45^o$ magnetic-field configurations.
The number of magnetic-reconnection sites increases with the ion-to-electron mass ratio and the Alfv\'enic Mach number. 
Runs with $\varphi=45^o$ demonstrate an almost twice larger number of magnetic reconnection sites and a slightly smaller electron density inside vortices than do simulations with $\varphi=0^o$, on account of substantial magnetic vortex production via magnetic turbulence in the shock ramp. \mpo{We do not observe a dependence on the plasma beta of the vortex number and the probability density function of electron density inside vortices.}

\item The growth rate of the Weibel instability at the shock foot is about of $\Gamma\approx0.1 \ompi$ in all simulations with in-plane magnetic field configuration. \mpo{The shock thickness in runs with mass ratio $\mi/\me \geq 100$ is sufficient to allow instability development into the nonlinear regime. The Weibel filaments become tearing-mode unstable on account of a temperature anisotropy, $T_{\perp}/T_{\parallel}>1$, and so they} decay through magnetic reconnection.

\item Interactions of electrons with magnetic reconnection sites lead to electron energization to nonthermal energies \mpo{through a number of mechanisms. We identify} acceleration in the $E_z$ electric field at an X-point (Speiser orbits), electron capture in the magnetic islands (``island surfing"), acceleration by bouncing between merging magnetic islands (first-order Fermi-like acceleration), and stochastic collisions with magnetic vortices (second-order Fermi-like acceleration).
The \mpo{development of magnetic reconnection in the shock ramp is truncated} by shock self-reformation that destroys magnetic filaments, and so reconnection reaches only the early stages of the process. 

\item Due to more frequent magnetic reconnection in runs with higher mass ratios and Alfv\'en Mach numbers, magnetic reconnection becomes the dominant \mpo{provider of nonthermal electrons downstream of the shock. As discussed in \cite{Bohdan2019b} the contribution of SSA is generally negligible.}


\end{itemize}

As mentioned, the magnetic reconnection becomes a dominant process for shocks with realistic physical parameters.
Simulations of quasi-perpendicular shocks exhibit shock-reflected electrons propagating upstream, where they can be responsible for production of magnetic turbulence. The electron reflection occurs at the shock foot/ramp, and magnetic reconnection may play a substantial role in this process. \ab{Therefore the role of magnetic reconnection in quasi-perpendicular shocks} should be clarified with further simulations.

\ab{The 3D simulations of \cite{Matsumoto2017} demonstrated the localized density clump in the \rev{foot} region, but they did not necessarily show signatures of magnetic reconnection.} 
Equation~\ref{Weibel_nonlin} is not satisfied in their study on account of the small mass ratio, and so the Weibel instability can not reach the nonlinear stage needed to trigger magnetic reconnection. \mpo{Generally, magnetic reconnection is expected to accelerate electrons more efficiently in 3D geometry than in the 2D case \citep{2015PhPl...22j0704D}.}
The scaling of magnetic-reconnection efficiency with mass ratio and $\ompe/\omce$ in 3D simulations should be similar to that in 2D case.

The issue of energy redistribution, electron heating processes, and the generation of turbulent magnetic field at perpendicular shocks in simulations will be covered in the \jn{forthcoming} last publication of this series.

\acknowledgments

The work of J.N. has been supported by Narodowe Centrum Nauki through research project DEC-2013/10/E/ST9/00662. This work was supported by JSPS-PAN Bilateral Joint Research Project Grant Number 180500000671.
This research was supported in part by PLGrid Infrastructure through a 10 Mcore-hour allocation on the 2.399 PFlop Prometheus system at ACC Cyfronet AGH. Part of the numerical work was conducted on resources provided by the North-German Supercomputing Alliance (HLRN) under projects bbp00003 and bbp00014.

\appendix

\section{The dispersion equation of ion Weibel instability for magnetized plasmas with perpendicular currents}\label{appendix}


\mpo{The dispersion relation of waves is usually calculated using a zeroth-order distribution function that satisfies the steady-state Vlasov equation and wave-like perturbations that satisfy the full Vlasov equation and Maxwell's equations \citep{1998JPlPh..60..111M}. In principle the analysis can be performed in any reference frame. In the presence of an ambient magnetic field, ${\bf B}_0$, however, the  plasma drift motion will cause a non-zero motional electric field, ${\bf E}_0$, that must be explicitely considered in the force term of the Vlasov equation.}

\mpo{The growth rate of the ion Weibel instability in our simulations is typically much smaller than the electron gyro-frequency, $\omce$, and so electrons will gyrate in, and co-move with, the large-scale magnetic field. In that case the electron rest frame is also the frame in which the motional electric field vanishes, in line with ideal magnetohydrodynamics (MHD) \citep{1965RvPP....1..205B}. The electron rest frame is therefore the preferred frame to calculate the dispersion relation of ion Weibel modes and used in a number of treatments \citep{1990PhRvL..65.1104C,1992PhFlB...4..719Y,1991PhFlB...3.3074Y,2001PlPhR..27..490S}.}

Let us consider a plasma whose components ($a=e,p$) have the following non-relativistic distribution function
\be 
f_{a}(v_x,v_y,v_z)= \frac{n_a}{\pi^{3/2}u_a^{3/2}} \exp\left( - \frac{(v_x-v_{a,x})^2 + (v_y-v_{a,y})^2 + v_z^2}{u_a^2} \right) ,
\label{distfunc}
\ee
where $u_a$ is the thermal speed, $n_a=\mathrm{const}$ the number density, and $v_{a,x}$ and $v_{a,y}$ are the drifting speeds in $x$ and $y$ directions, respectively. We note that 
\jn{some particle populations may be composed}
of several components with different temperatures and streaming velocities. The ambient magnetic field, ${\bf B}_0$, is directed along the $z$ axis. 
 
\mpo{The dispersion equation for ion Weibel modes with wave vectors parallel to the ambient magnetic field (${\bf k}=(0,0,k)$) reads: }
\be 
\Lambda= \begin{vmatrix}
\Lambda_{xx} & \Lambda_{xy} & \Lambda_{xz} \\
\Lambda_{yx} & \Lambda_{yy} & \Lambda_{yz} \\
\Lambda_{zx} & \Lambda_{zy} & \Lambda_{zz} 
\end{vmatrix}=0 ,
\label{ap0}
\ee
where the matrix elements are given by 
\be 
\Lambda_{ij}= \epsilon_{ij}(\omega,{\bf k}) - \l(kc\over \omega \r)^2\delta_{ij} + {k_i k_j c^2\over\omega^2},
\label{ap1}
\ee
where $\delta_{ij}$ is the Kronecker symbol and 
\be
\epsilon_{xz}=\epsilon_{zx}= - \sum_a \frac{\omega_{p,a}^2v_{a,x}}{\omega ku_a^2}Z'\l(\frac{\omega}{ku_a}\r),
\label{ap2}
\ee

\be
\epsilon_{yz}=\epsilon_{zy}= - \sum_a \frac{\omega_{p,a}^2v_{a,y}}{\omega ku_a^2}Z'\l(\frac{\omega}{ku_a}\r),
\label{ap3}
\ee

\be
\epsilon_{zz}=1 - \sum_a \frac{\omega_{p,a}^2}{ (ku_a)^2}Z'\l(\frac{\omega}{ku_a}\r),
\label{ap4}
\ee


\begin{multline}
\epsilon_{xx}=1 - \sum_a \frac{\omega_{p,a}^2}{2\omega^2 }\l[ 
Z'\l(\frac{\omega-\Omega_a}{ku_a}\r)\l( \frac{1}{2}+\l(\frac{v_{a,x}}{u_a}\r)^2+i\frac{v_{a,x}v_{a,y}}{u_a^2}\r)
+Z'\l(\frac{\omega+\Omega_a}{ku_a}\r)\l( \frac{1}{2}+\l(\frac{v_{a,x}}{u_a}\r)^2-i\frac{v_{a,x}v_{a,y}}{u_a^2}\r)\r. \\ \l.
+ 2 + \frac{\Omega_a}{ku_a}\l( Z\l(\frac{\omega+\Omega_a}{ku_a}\r) - Z\l(\frac{\omega-\Omega_a}{ku_a}\r)\r)   
\r],
\label{ap5}
\end{multline}

\begin{multline}
\epsilon_{yy}=1 - \sum_a \frac{\omega_{p,a}^2}{2\omega^2 }\l[ 
Z'\l(\frac{\omega-\Omega_a}{ku_a}\r)\l( \frac{1}{2}+\l(\frac{v_{a,y}}{u_a}\r)^2+i\frac{v_{a,x}v_{a,y}}{u_a^2}\r)
+Z'\l(\frac{\omega+\Omega_a}{ku_a}\r)\l( \frac{1}{2}+\l(\frac{v_{a,y}}{u_a}\r)^2-i\frac{v_{a,x}v_{a,y}}{u_a^2}\r)\r. \\ \l.
+ 2 + \frac{\Omega_a}{ku_a}\l( Z\l(\frac{\omega+\Omega_a}{ku_a}\r) - Z\l(\frac{\omega-\Omega_a}{ku_a}\r)\r)   
\r],
\label{ap6}
\end{multline}

\begin{multline}
\epsilon_{xy}= i\sum_a \frac{\omega_{p,a}^2}{2\omega^2 }\l[  
iZ'\l(\frac{\omega-\Omega_a}{ku_a}\r)\l( \frac{v_{a,x}v_{a,y}}{u_a^2}+\frac{i}{2}+i\l(\frac{v_{a,y}}{u_a}\r)^2\r)
+iZ'\l(\frac{\omega+\Omega_a}{ku_a}\r)\l( \frac{v_{a,x}v_{a,y}}{u_a^2}-\frac{i}{2}-i\l(\frac{v_{a,y}}{u_a}\r)^2\r)\r. \\ \l.
+  \frac{\Omega_a}{ku_a}\l( Z\l(\frac{\omega+\Omega_a}{ku_a}\r) + Z\l(\frac{\omega-\Omega_a}{ku_a}\r)\r)   
\r],
\label{ap7}
\end{multline}

\begin{multline}
\epsilon_{yx}= i\sum_a \frac{\omega_{p,a}^2}{2\omega^2 }\l[  
Z'\l(\frac{\omega-\Omega_a}{ku_a}\r)\l( \frac{1}{2}+\l(\frac{v_{a,x}}{u_a}+i\frac{v_{a,x}v_{a,y}}{u_a^2}\r)^2\r)
-Z'\l(\frac{\omega+\Omega_a}{ku_a}\r)\l( \frac{1}{2}+\l(\frac{v_{a,x}}{u_a}\r)^2-i\frac{v_{a,x}v_{a,y}}{u_a^2}\r)\r. \\ \l.
-  \frac{\Omega_a}{ku_a}\l( Z\l(\frac{\omega+\Omega_a}{ku_a}\r) + Z\l(\frac{\omega-\Omega_a}{ku_a}\r)\r)   
\r].
\label{ap8}
\end{multline}
Here, $\Omega_{a}$ is the gyro-frequency, $\omega_{p,a}$ is the plasma frequency, and $Z'(x)$ the plasma dispersion function \citep{1961pdf..book.....F}. 

\mpo{In their simulation of a collisionless shock \citet{2010ApJ...721..828K} observed plasma filaments that they identified with ion Weibel modes.
We repeated their calculation and found matching results.}
Note that in \cite{2010ApJ...721..828K} plasma have been \jn{split} into 4 beams, and we verified that one finds to within 10\% the same growth rate when using three beams, namely electrons, cold upstream ions, and hot reflected ions. 

\begin{figure}[htb]
\centering
\includegraphics[width=0.47\linewidth]{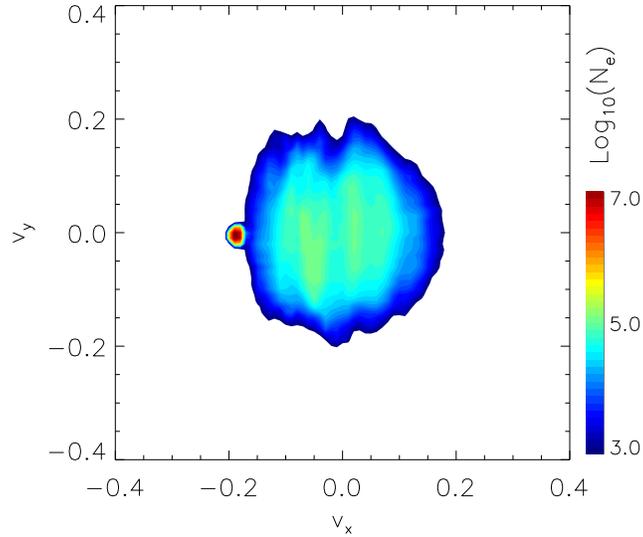}
\caption{Phase-space density of ions at the shock foot (taken from run F2). }
\label{growthrates}
\end{figure}

A phase-space plot of ions extracted from the shock foot is presented in Fig~\ref{growthrates}. The resulting growth rate, $\Gamma(k)$, is shown in Fig.~\ref{growsrate} in the main text.


\bibliographystyle{apj}
\bibliography{main}

\end{document}